# High-precision temporal interferometry from nonlinearly-spaced phase-shifted interferograms through spatial-filtering

MANUEL SERVIN,* MOISES PADILLA, GUILLERMO GARNICA, AND GONZALO PAEZ

*Centro de Investigaciones en Optica A. C., Loma del Bosque 115, 37150 Leon Guanajuato, Mexico.*
*\*mservin@cio.mx*

**Abstract**: We present a high-precision temporal-spatial phase-demodulation algorithm for phase-shifting interferometry (PSI) affected by random/systematic phase-stepping errors. Laser interferometers in standard optical-shops suffer from several error sources including random phase-shift deviations. Even calibrated phase-shifters do not achieve floating-point linear accuracy, as routinely obtained in multimedia video-projectors for fringe-projection profilometry. In standard optical-shops, calibrated phase-shifting interferometers suffer from nonlinearities due to vibrations, turbulence, and environmental fluctuations (temperature, pressure, humidity, air composition) still under controlled laboratory conditions. These random phase-step errors (even if they are small), increases the uncertainty of the phase measurement. This is particularly significant if the wavefront tolerance is tightened to high precision optics. We show that these phase-step errors precludes high-precision wavefront measurements because its uncertainty increases to around $\lambda/10$. We develop an analytical expression based on optical-wavefront formalism showing that these phase-step nonlinearities appear as a spurious conjugate signal degrading the desired wavefront. Removing this spurious conjugate constitutes the central objective of the proposed nonlinear phase-shifting algorithm (nPSA). Using this nPSI algorithm we demodulate experimental interferograms subject to small vibrations and phase-shifter nonlinearities, obtaining a high-precision spurious-free, demodulated wavefront. We show that our artifact-free, temporal-spatial quadrature filtering, accomplishes an equivalent wavefront precision as the one obtained from floating-point linear phase-shifting interferometry.

## 1. Introduction

Temporal phase-shifting interferometry (PSI) is a powerful and well established technique to measure wavefronts with high precision [1-6]. The first paper on temporal phase-shifting interferometry (PSI) was Carré [2]. This 1966 paper was in some ways years ahead of its time, but it was un-noticed because its application was not to two-dimensional (2D) PSI. The first linear *N*-step phase-shifting algorithm (PSA) was due to Bruning et al. [3]. Bruning et al. mention several times, that avoiding systematic/random phase-step errors, and averaging many fringe patterns, one would be able to attain $\lambda/100$ wavefront accuracy [3]. In the absence of systematic phase-step errors, temporal averaging reduces the wavefront noise power as (1/*N*); one would need 50 fringe samples to attain $\lambda/100$ accuracy [1-6]. As mentioned, systematic phase-shifting errors rarely reduce by temporal averaging. For all these reasons, typical optical-shops (ours in particular) rarely obtain better than $\lambda/10$ wavefront accuracy. Here we show that PSI with few fringes interferograms, one cannot spot an artifact phase-error in the demodulated wavefront. That is because the low-frequency artifact "hides away" within the measuring phase. To this day, research in PSAs reveals that random-systematic phase-shifting errors build-up and $\lambda/100$ precision are almost never obtained; air-turbulence, vibrations and small phase-shifting nonlinearities being difficult to control [1-41]. Schmitz et al., at the US National Institute of Standards and Technology (NIST) have reported how sensitive phase-shifting interferometers are to mechanical-environmental conditions which translates into uncertainties and repeatability demodulated wavefront artifacts [41]. In most optical-shop testing facilities some of these error sources translates into random-systematic phase-shifting nonlinearities.

The first works on nonuniform phase-shifting algorithms (nPSA) may be seen in references [7-11]. Least-squares nPSA (LS-nPSA) were proposed by Morgan [7], and Greivenkamp [8]. Afterwards came the generalized/iterative least-squares gLS-nPSA [9-12]. In gLS-nPSA the demodulated phase and nonlinear phase-steps are iteratively estimated. In this way the global nonlinear phase-estimation problem is broken into two iterated linear systems converging (with some remaining error) to the searched phase, and the nonlinear phase-steps [12-23]. A more recent variation gLS-nPSA, named advanced iterative algorithm (AIA), became popular [13]. Depending on the phase-step number and fringe-noise, no better than $\lambda/20$ tolerances are obtained by gLS-nPSA.

Another alternative to nPSI is to use the Lissajous figure of the demodulated complex-valued wavefront [24-28]. The Lissajous figure is the parametric plot of the real and imaginary parts of the demodulated signal. A Lissajous ellipse is the hallmark of an erroneous wavefront demodulation while a Lissajous circle is synonymous of good demodulation [24-28]. As in gLS-nPSA, the noise of the complex-valued estimated wavefront limit the least-squares fitting of the Lissajous ellipse [24-28]. Depending on the amount of systematic-random phase-step errors, wavefront tolerances around $(\lambda/20)$ may be attained. That is because the least-squares fit to a noisy Lissajous ellipse is pretty sensitive to noise.

In 2011 Vargas et al. [29] used a statistical technique called principal component analysis (PCA) for nonuniform phase-shifting interferometry (nPSI) [29]. We call this procedure the PCA-nPSI algorithm. The PCA algorithm was published in 1901 by Karl Parson [30]. Pearson used PCA to find few orthogonal (uncorrelated) signals from a very large set of correlated statistical data [30]. The PCA applied to nPSI (PCA-nPSI) estimates the sine and cosine of the estimated phase from nonuniform phase-shifted fringes [29]. In other words, the linear PCA-nPSI algorithm simultaneously estimate the modulating phase and the nonlinear phase-steps of the interferograms [29-38]. The PCA-nPSI algorithm being an optimum linear system, cannot accurately solve the nonlinear nPSI problem. That is why it is not surprising that PCA-nPSI has serious problems that have been studied since its introduction [31-38]. Moreover, Karl Pearson proposed the PCA as a statistical analysis to find few principal components from a very large set of statistical data [30]; not as a nPSI algorithm depending on a handful of samples. So in general the PCA-nPSI would give a fairly bad estimate to the modulating wavefront. The PCA-nPSI being linear give in general, lower wavefront precision than gLS-nPSA algorithms [12-23]. Therefore better PCA-nPSI algorithms have been published using the gLS-PSI as final step [31-37]. Some other improvements rely on applying the Lissajous figure to PCA-nPSI, taking as first approximation the PCA-nPSI estimation. In brief the PCA-nPSI is not in general a reliable technique to estimate the wavefront from nonuniform phase-stepped fringes.

Linear phase-shifting interferometry may use the frequency transfer function (FTF) to analyze the properties of temporal quadrature-filters in the Fourier domain [1]. Knowing the FTF one can easily find the signal-to-noise ratio (SNR) and harmonic sensitivity of PSAs [1]. Knowing the nonlinear phase-steps, we recently published a nPSI algorithm that use the desired FTF's spectral zeroes to find the coefficients of the nPSA [39]. Knowledge of the FTF of a nPSA [39] allows one to calculate the SNR and fringe harmonics sensitivity, as routinely done for linear PSAs [1].

All nPSI algorithms (except PCA-nPSI) estimate the modulating-phase and nonlinear phase-steps iteratively. In nPSI the wavefront demodulation error (except for the data noise) comes from the limited accuracy of the nonuniform phase-steps estimation [7-38]. Given that avoiding the fringe noise is impossible, here we bypass the nonlinear phase-steps estimation process. This is possible, if and only if, one can introduce spatial-carrier fringes to the temporal interferograms. As we shown, the proposed nPSA entirely bypasses the nonlinear phase-steps estimation. In this way the demodulated wavefront accuracy do not depend on the phase-step estimation precision, obtaining an estimated wavefront as reliable as the one obtained by an ideal floating-point linear, phase-shifter. Here the experimental interferograms were digitized by an upgraded WYKO-6000 Fizeau interferometer with calibrated PZT phase-shifter. In spite that the WYKO-6000 is on top of a Newport optical-table it is still sensitive to vibrations and PZT small nonlinear deviations. The kind of demodulation wavefront artifacts herein described is a daily trouble in our optical-shop facility, and we had to solve it once and for all. The proposed nPSI algorithm is new, accurate, straightforward, and certainly useful in high-quality optical manufacturing shops worldwide.

## 2. Two formalisms for digital phase-shifting interferometry (PSI)

The temporal interferograms degraded by nonlinear phase-shifting errors are modeled by,

$$I(n) = a + b\cos[\varphi + \omega_0 n + \varepsilon_n]; \quad n \in \{0,...,N-1\}. \qquad (1)$$

The interferograms are $I(n) = I(x,y,n)$; the modulating phase is $\varphi = \varphi(x,y)$; the wavefront is $W(x,y) = (\lambda/2\pi)\varphi(x,y)$; being $\lambda$ the laser wavelength. The background is $a = a(x,y)$ and the contrast is $b = b(x,y)$. The linear phase-steps are $\omega_0 n$; the nonlinear deviations are $\{\varepsilon_n\}$ assumed space-independent [6]. The most common $N$-steps PSI formalisms are,

$$\tan(\hat{\varphi}) = \frac{\sum_{n=0}^{N-1} c_n \sin(n\omega_0) I(n)}{\sum_{n=0}^{N-1} c_n \cos(n\omega_0) I(n)}, \quad Ae^{i\hat{\varphi}} = \sum_{n=0}^{N-1} c_n e^{in\omega_0} I(n). \qquad (2)$$

Being $\hat{\varphi} = \hat{\varphi}(x,y)$ the estimated phase, and usually $c_n \in \mathbb{R}$ [1]. The $\tan(\hat{\varphi})$ formula is used since 1974 [3,40], and more recently the optical-wavefront formalism $Ae^{i(2\pi/\lambda)\hat{W}} = Ae^{i\hat{\varphi}}$ [1]. Both formulas give the same phase estimation $\hat{\varphi} = \hat{\varphi}(x,y)$.

## 3. Phase-shifting interferometry with nonlinear phase-step errors (nPSI)

Here we present the mathematical theory of our proposed temporal-spatial nPSI algorithm.

### 3.1 Phase-shifting interferometry with nonlinear phase-step errors

Using the fringe model $I(n) = I(x,y,n)$ in Eq. (1), the $\tan(\hat{\varphi})$ formalism reads,

$$\tan(\hat{\varphi}) = \frac{\sum_{n=0}^{N-1} c_n \sin(n\omega_0)\left[a + b\cos(\varphi + \omega_0 n + \varepsilon_n)\right]}{\sum_{n=0}^{N-1} c_n \cos(n\omega_0)\left[a + b\cos(\varphi + \omega_0 n + \varepsilon_n)\right]}; \quad \forall(x,y). \qquad (3)$$

Due to the tangent nonlinearity, the mathematically proofs for analyzing even elementary properties of linear PSAs are cumbersome, taking many steps of algebra and trigonometry [40]. And usually, only approximations are possible, take for example the analysis for linear-detuning error [40]. As a consequence, by looking at $\tan(\hat{\varphi})$ one cannot spot the reason why phase-step nonlinearities $\{\varepsilon_n \neq 0\}$ would give an erroneous phase $\hat{\varphi}(x,y)$; except for the circular-reasoning that these nonlinearities must play a role.

On the other hand, the analytic-signal $Ae^{i\hat{\varphi}}$ formalism is in general, more effective for analyzing PSI algorithms [1]. Using the fringes $I(n)$, the wavefront formalism gives,

$$Ae^{i\hat{\varphi}} = \sum_{n=0}^{N-1} c_n e^{-in\omega_0} I(n) = \sum_{n=0}^{N-1} c_n e^{-in\omega_0}\left\{a + \frac{b}{2}e^{i[\varphi+n\omega_0+\varepsilon_n]} + \frac{b}{2}e^{-i[\varphi+n\omega_0+\varepsilon_n]}\right\}. \quad (4)$$

Obtaining,

$$Ae^{i\hat{\varphi}} = a\sum_{n=0}^{N-1} c_n e^{-in\omega_0} + \frac{b}{2}\sum_{n=0}^{N-1} c_n e^{i[\varphi+\varepsilon_n]} + \frac{b}{2}\sum_{n=0}^{N-1} c_n e^{-i[\varphi+2n\omega_0+\varepsilon_n]}. \quad (5)$$

All well designed PSI algorithms rejects the background, $a\sum c_n e^{-in\omega_0} = 0$,

$$Ae^{i\hat{\varphi}} = A_1 e^{i\varphi} + A_2 e^{-i\varphi}; \qquad |A_1| > |A_2|,$$
$$A_1 = \frac{b}{2}\sum_{n=0}^{N-1} c_n e^{i\varepsilon_n}, \quad A_2 = \frac{b}{2}\sum_{n=0}^{N-1} c_n e^{-i[2n\omega_0+\varepsilon_n]} \;; \forall(x,y) \quad (6)$$

Here the estimated signal is $Ae^{i\hat{\varphi}}$; being $A_1 e^{i(2\pi/\lambda)W} = A_1 e^{i\varphi}$ the searched wavefront, while its conjugate $A_2 e^{-i(2\pi/\lambda)W} = A_2 e^{-i\varphi}$ is a spurious/artifact signal. The demodulated phase $\hat{\varphi}(x,y) = \arg[A_1 e^{i\varphi} + A_2 e^{-i\varphi}]$ has a phase artifact with double-frequency fringe structure, similar to detuning in linear PSAs [1,40] (see Fig. 1) [1,40]. Equation (6) shows that nonlinear errors $\{\varepsilon_n\}$ pop-up as a spurious conjugate $A_2 e^{-i\varphi}$. As far as we know, Eq. (6) is a new and useful result for better understanding PSAs degraded by nonlinear phase-steps.

### 3.2 Simulation of the artifact phase error due to phase-shifter nonlinearities

Here we simulate Eq. (6). Figure 1 shows the phase demodulation-error given by,

$$\varphi_{Error} = \varphi - \arg\left[e^{i\varphi} + 0.1 e^{-i\varphi}\right]; \quad \forall(x,y). \quad (7)$$

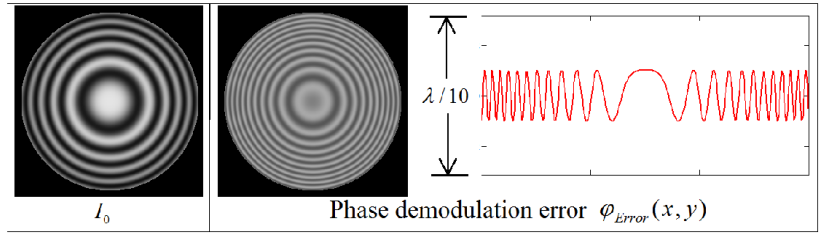

Fig. 1 Simulation of the artifact phase-error due to small phase-steps nonlinearities. At left, a sampled interferogram, and at right, the double-frequency fringe phase-error in Eq. (7).

The phase-demodulation artifact shown in Fig. 1 is the hallmark of small phase-step nonlinear deviations. Note also that this spurious-structure looks like detuning error in linear PSI [1,40].

### 3.3 Spatial-carrier for filtering-out the spurious conjugate wavefront

Introducing a spatial-carrier, the searched phase changes to $[\varphi(x,y) + u_0 x]$,

$$I(n) = a + b\cos[\varphi + u_0 x + \omega_0 n + \varepsilon_n]; \quad u_0 > \left|\frac{\partial \varphi}{\partial x}\right|_{Max} ; \forall(x,y). \quad (8)$$

Substituting $\varphi(x,y) \to [\varphi(x,y) + u_0 x]$ in Eq. (6) one obtains,

$$Ae^{i\hat{\varphi}} = A_1 e^{i[\varphi+u_0 x]} + A_2 e^{-i[\varphi+u_0 x]}; \qquad |A_1| > |A_2|. \quad (9)$$

The signals $A_1$, $A_2$ are still given by Eq. (6). But now in Eq. (9) the spurious $A_2 e^{-i[\varphi+u_0 x]}$ has a different spectral location that can be eliminated by spatial low-pass filtering $LPF[\cdot]$ as,

$$LPF\left[\left(A e^{i\hat{\varphi}}\right) e^{-i u_0 x}\right] = \left(\frac{b}{2}\sum_{n=0}^{N-1} c_n e^{i\varepsilon_n}\right) e^{i\varphi} = A_1 e^{i\varphi}. \tag{10}$$

Thus the artifact-free estimated phase is,

$$\tan[\hat{\varphi}] = \frac{A_1 \sin[\varphi]}{A_1 \cos[\varphi]} = \frac{\sin[\varphi]}{\cos[\varphi]}; \quad A_1 \neq 0. \tag{11}$$

The signal $A_1 = (b/2)\sum c_n e^{i\varepsilon_n}$ disappears regardless of $\{\varepsilon_n\}$. The estimation of $\hat{\varphi}(x,y)$ no longer depends in anyway of the nonlinearities $\{\varepsilon_n\}$. Thus the estimated phase in Eq. (11) is as accurate as the one obtained using an ideal floating-point linear phase-shifter.

## 4. Experimental nonuniform phase-shifted Interferograms demodulation

We used a WYKO-6000 Fizeau interferometer and the 5-steps Schwider-Hariharan (SH-PSI) algorithm ($\omega_0 = 2\pi/4$) [1,40]. The FTF ($H_{SH}(\omega)$) and $A e^{i\hat{\varphi}}$ formula are given by [1],

$$\begin{aligned} H_{SH}(\omega) &= \left(1-e^{i\omega}\right)\left[1-e^{i(\omega+\omega_0)}\right]^2 \left[1-e^{i(\omega+2\omega_0)}\right], \\ A e^{i\hat{\varphi}} &= I(0) + 2i\,I(1) - 2\,I(2) - 2i\,I(3) + I(4); \quad \forall (x,y). \end{aligned} \tag{12}$$

The $|H_{SH}(\omega)|$ plot, with normalized frequency is shown in Fig. 2.

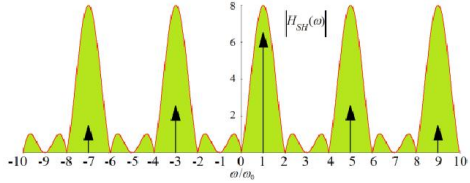

Fig. 2. The $|H_{SH}(\omega)|$ plot. The SH-PSA is robust to linear-detuning, and pass the harmonics at (-7, -3, 5, 9) in the frequency range shown.

The SH-PSA robust to linear detuning, is however sensitive to nonlinear phase-shifting errors.

### 4.1 Experimental phase demodulation with the 5-step, SH-PSI algorithm

Figure 3 shows the upgraded WYKO-6000 Fizeau interferometer used for the experiments.

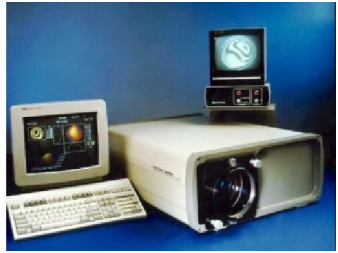

Fig. 3. WYKO-6000 Fizeau interferometer. The B&W video-monitor at top, allows real-time fringe visualization. The computer, the CCD-camera, and fringe-software were upgraded.

Using the WYKO's we took the phase-shifted fringes (with $\omega_0 = \pi/2$) shown in Fig. 4.

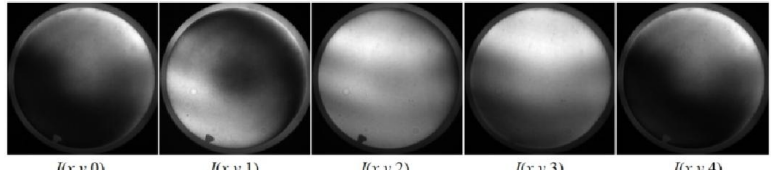

Fig. 4. Five interferograms of a manufacturing flat with nominal phase-step of ($\pi/2$). The WYKO's video-monitor show randomly vibrating fringes despite the stabilizing optical-table.

We have demodulated 50 phases in a row, randomly picking $\{\hat{\varphi}_1, \hat{\varphi}_2, \hat{\varphi}_3\}$ shown in Fig. 5. At first sight, the phases $\{\hat{\varphi}_1, \hat{\varphi}_2, \hat{\varphi}_3\}$ look identical, but they are different. The right hand side panel showing $\{10(\hat{\varphi}_2 - \hat{\varphi}_1), 10(\hat{\varphi}_3 - \hat{\varphi}_1), 10(\hat{\varphi}_2 - \hat{\varphi}_3)\}$ expose the artifact phase-error.

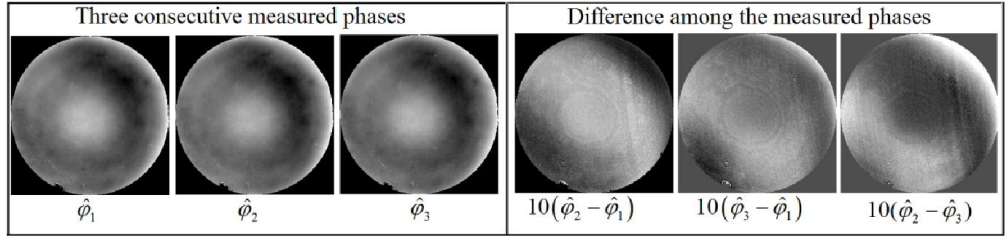

Fig. 5. Three phase measurements using the WYKO and SH-PSA (black-to-white maps to (0-$2\pi$) radians). At left panel all 3 phases look identical, but they have ($\lambda/10$) peak phase-difference among them, this is shown at the right hand side panel.

The differences $\{10(\hat{\varphi}_2 - \hat{\varphi}_1), 10(\hat{\varphi}_3 - \hat{\varphi}_1), 10(\hat{\varphi}_2 - \hat{\varphi}_3)\}$, show a ($\lambda/10$) peak-error artifact. If the required tolerance is ($\lambda/10$), then any $\{\hat{\varphi}_1, \hat{\varphi}_2, \hat{\varphi}_3\}$ is equally useful. Thus, vibrations and systematic phase-shifter errors (however small) introduces ($\lambda/10$) repeatability-reliability artifacts. Note that the low spatial-frequency of the phase-artifact makes impossible to spot a good measurement among many phase estimations.

*4.2 Phase estimation with the proposed temporal-spatial algorithm*

Figure 6 shows five WYKO's spatial-carrier interferograms with nominal $\omega_0 = \pi/2$.

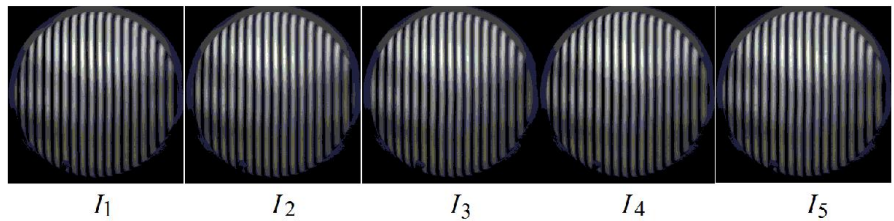

Fig. 6. Five digitized WYKO spatial-carrier interferograms with nominal phase-step of ($\pi/2$) radians/sample. By looking at the WYKO's video-monitor, the fringes vibrate randomly, with peak deviation roughly of ($\lambda/10$); in spite of the optical-table,

In Fig. 6 the fringes have $\omega_0 = \pi/2$ nominal phase-step. The WYKO's video-monitor show that the fringes vibrate roughly ($\lambda/10$); so random/systematic peak-errors $\{\varepsilon_n\} \sim (\lambda/10)$ are expected. From interferograms like those shown in Fig. 6 we demodulated five phases $\{\hat{\varphi}_1, \hat{\varphi}_2, \hat{\varphi}_3, \hat{\varphi}_4, \hat{\varphi}_5\}$ using the 5-step SH-PSA [1].

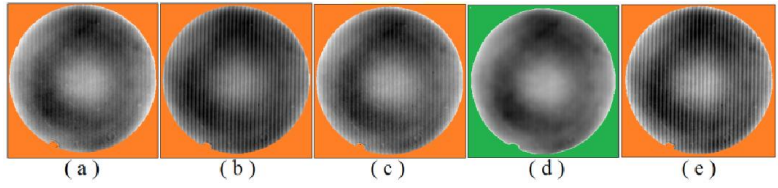

Fig. 7. Five consecutive wavefronts of the same optical-flat using SH-PSA. Except for the green-panel all others show double-frequency fringe-structure, an artifact phase-error.

Figure 7 clearly shows the advantage of using temporal-spatial carrier. Any systematic-random nonlinear phase-step error is easily spot as a double-frequency fringe-artifact. The green-panel in 7(d) has no visible phase-artifact; it has no phase error. In our optical-shop, an artifact-free phase (panel (d)) occurs in about 1 out-of 50 phase estimations.

*4.3 Phase estimation with the proposed temporal-spatial demodulation*

Figure 8 shows the spectrum of $[A_1 e^{i[\varphi+u_0 x]} + A_2 e^{-i[\varphi+u_0 x]}]$, and $\hat{\varphi} = \arg[A_1 e^{i\varphi} + A_2 e^{-i[\varphi+2u_0 x]}]$.

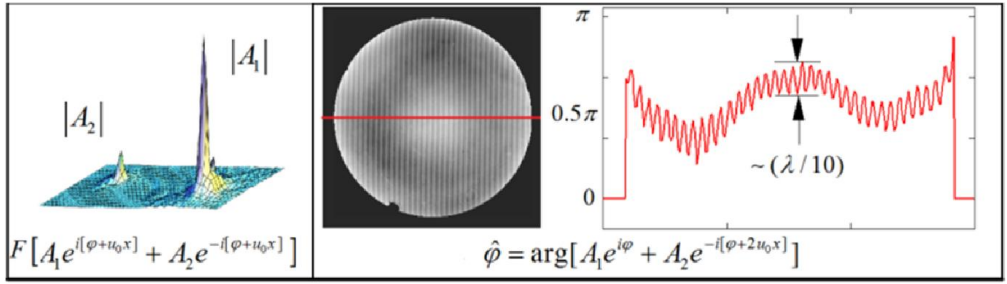

Fig. 8. The spectrum of $[A_1 e^{i[\varphi+u_0 x]} + A_2 e^{-i[\varphi+u_0 x]}]$, and the phase $\hat{\varphi} = \arg[A_1 e^{i\varphi} + A_2 e^{-i[\varphi+2u_0 x]}]$. The red-line shows the double-frequency fringe phase-error artifact.

In Fig. 8, the spectrum of $[A_1 e^{i[\varphi+u_0 x]} + A_2 e^{-i[\varphi+u_0 x]}]$ shows the small-amplitude of $A_2 e^{-i[\varphi+u_0 x]}$.

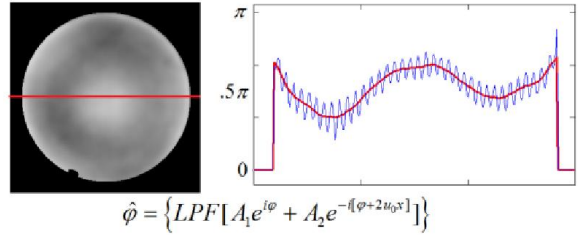

Fig. 9. The estimated phase $\hat{\varphi} = \hat{\varphi}(x,y)$. The red-plot from the red-line-cut show the artifact-free demodulated phase. The filtered-out phase-artifact is shown in blue.

Finally Fig. 9 shows the spatial low-pass filtered wavefront $Ae^{i\hat{\varphi}} = LPF[A_1 e^{i\varphi} + A_2 e^{-i[\varphi+2u_0 x]}]$ with no phase-error artifact.

*4.4 Advantages and limitations of the proposed temporal-spatial nPSI algorithm*

Due to interferogram's noise, published nPSI algorithms cannot attain floating-point phase-step estimation accuracy. This fact precludes a conjugate-free wavefront estimation (see Eqs. (6)-(9)). We think that the only way of having better than $\lambda/10$ accuracy from nonlinearly-spaced phase-step fringes is to bypass the phase-step estimation entirely. Therefore, instead of estimating the nonuniform phase-steps, we propose to introduce spatial-carrier to the phase-shifted interferograms. We finally use spatial-filtering to bypass the phase-step nonlinearities and obtain a conjugate-free demodulated wavefront. The sole limitation to this temporal-spatial nPSI algorithm is to introduce spatial-carrier. In the rare cases were this is not possible, the proposed nPSI technique cannot be used.

## 5. Summary

We have presented a temporal-spatial nonuniform phase-shifting interferometry (nPSI) algorithm to accurately demodulate temporal interferograms having phase-step errors. The nonlinear phase-step errors $\{\varepsilon_n\}$ may arise from air-turbulence, vibrations and phase-shifter nonlinearities [41]. We used a WYKO-6000 Fizeau interferometer located on top of a Newport optical table. In spite of this, the WYKO's video-monitor shows visible fringe vibrations of roughly ($\lambda/10$) peak amplitude. Also we have found that the WYKO's PZT, has systematic and small phase-step nonlinearities. Adding up these experimental phase-step errors the demodulated wavefront is usually within $\lambda/10$ tolerance. Here we have shown typical WYKO interferograms with nominal phase-shifts of ($\pi/2$). Even with small phase-step deviations $\{\varepsilon_n\}$, the demodulated phase have notorious double-frequency-fringe artifact.

For phase-demodulation, we used the 5-step, Schwider-Hariharan phase-shifting algorithm (SH-PSA) which is robust to linear-detuning [1]. The SH-PSA is however very sensitive to nonlinear phase-step deviations $\{\varepsilon_n\}$. This sensitivity give rise to a spurious-conjugate wavefront which degrades the wavefront estimation (Eqs. (6)-(9)). We finally use carrier-frequency interferograms for spatially filtering-out the spurious-conjugate. In this way one finally obtains an artifact-free demodulated wavefront. Put in other words, applying the proposed temporal-spatial technique, one obtains an estimated wavefront as reliable as if we were using a floating-point accurate, linear phase-shifter.

## References

1. M. Servín, J. A Quiroga, and M. Padilla, *Fringe Pattern Analysis for Optical Metrology*, (WILEY-VCH, 2014).
2. P. Carré, " Installation et utilisation du comparateur photoelectrique et interferentiel du Bureau International des Poids et Mesures," Metrologia **2**, 13-22 (1966).
3. J. H. Bruning, D. R. Herriott, J. E. Gallagher, D. P. Rosenfeld, A. D. White, and D. J. Brangaccio, "Digital Wavefront Measuring Interferometer for Testing Optical Surfaces and Lenses," Appl. Opt. **13**, 2693-2703 (1974).
4. P. Hariharan, B. F. Oreb, and T. Eiju, "Digital phase-shifting interferometry: a simple error-compensating phase calculation algorithm," Appl. Opt. **28**, 2504-2505 (1987).
5. K. Kinnstaetter, A. W. Lohmann, J. Schwider, and N. Streibl, "Accuracy of phase shifting interferometry," Appl. Opt. **24**, 5082-5089 (1988).
6. K. Creath, "Phase measurement interferometry: BEWARE these errors," Proc. SPIE **1553**, 213-220 (1991).
7. C. J. Morgan, "Least-squares estimation in phase-measurement interferometry," Opt. Lett. 7, 368-370 (1982).
8. J. E. Greivenkamp, "Generalized data reduction for heterodyne interferometry," Opt. Eng. 23, 350-352 (1984)
9. G. Lai and T. Yatagai, "Generalized phase-shifting interferometry," J. Opt. Soc. Am. A **8**, 822–827 (1991).
10. K. Okada, A. Sato, and J. Tsujiuchi, "Simultaneous calculation of phase distribution and scanning phase shift in phase shifting interferometry," Opt. Commun. **84**, 118–124 (1991).
11. C. T. Farrell and M. A. Player, "Phase step measurement and variable step algorithms in phase-shifting interferometry," Meas. Sci. Technol. **3**, 953–958 (1992).
12. I-B. Kong and S.-W. Kim, "General algorithm of phase-shifting interferometry by iterative least-squares fitting," Opt. Eng. **34**, 183–188 (1995).
13. Z. Wang and B. Han, "Advanced iterative algorithm for phase extraction of randomly phase-shifted interferograms," Opt. Lett. **29**, 1671–1673 (2004).
14. A. Patil and P. Rastogi, "Approaches in generalized phase shifting interferometry," Opt. Lasers Eng. **43**, 475–490 (2005).
15. H. Guo, Y. Yu, and M. Chen, "Blind phase shift estimation in phase shifting interferometry," J. Opt. Soc. Am. A **24**, 25–33 (2007).
16. X. F. Xu, L. Z. Cai, Y. R. Wang, X. F. Meng, W. J. Sun, H. Zhang, X. C. Cheng, G. Y. Dong, and X. X. Shen, "Simple direct extraction of unknown phase shift and wavefront reconstruction in generalized phase-shifting interferometry: algorithm and experiments," Opt. Lett. **33**, 776–778 (2008).
17. P. Gao, B. Yao, N. Lindlein, K. Mantel, I. Harder, and E. Geist, "Phase shift extraction for generalized phase-shifting interferometry," Opt. Lett. **34**, 3553–3555 (2009).
18. J. Deng, H. Wang, D. Zhang, L. Zhong, J. Fan, and X. Lu, "Phase shift extraction algorithm based on Euclidean matrix norm," Opt. Lett. **38**, 1506–1508 (2013).

19. H. Guo and Z. Zhang, "Phase shift estimation from variances of fringe pattern differences," Appl. Opt. **52**, 6572–6578 (2013).
20. R. Juarez-Salazar, C. Robledo-Sánchez, C. Meneses-Fabian, F. Guerrero-Sánchez, and L. A. Aguilar, "Generalized phase-shifting interferometry by parameter estimation with the least squares method," Opt. Lasers Eng. **51**, 626–632 (2013).
21. J. Deng, L. Zhong, H. Wang, H. Wang, W. Zhang, F. Zhang, S. Ma, and X. Lu, "1-Norm character of phase shifting interferograms and its application in phase shift extraction," Opt. Commun. **316**,156–160 (2014).
22. Q. Liu, Y. Wang, J. He, and F. Ji, "Phase shift extraction and wavefront retrieval from interferograms with background and contrast fluctuations," J. Opt. **17**, 025704 (2015).
23. Y. Xu, Y. Wang, Y. Ji, H. Han, and W. Jin, "Three-frame generalized phase-shifting interferometry by a Euclidean matrix norm algorithm," Opt. Lasers Eng. **84**, 89–95 (2016).
24. A. Albertazzi Jr., A. V. Fantin, D. P. Willemann, and M. E. Benedet, "Phase maps retrieval from sequences of phase shifted images with unknown phase steps using generalized N-dimensional Lissajous figures—principles and applications," Int. J. Optomechatron. **8**, 340–356 (2014).
25. C. Meneses-Fabian and F. A. Lara-Cortes, "Phase retrieval by Euclidean distance in self-calibrating generalized phase-shifting interferometry of three steps," Opt. Express **23**, 13589–13604 (2015).
26. F. Liu, Y. Wu, and F. Wu, "Correction of phase extraction error in phase-shifting interferometry based on Lissajous figure and ellipse fitting technology," Opt. Express **23**, 10794–10807 (2015).
27. F. Liu, Y. Wu, F. Wu, and W. Song, "Generalized phase shifting interferometry based on Lissajous calibration technology," Opt. Lasers Eng. **83**, 106–115 (2016).
28. F. Liu, J. Wang, Y. Wu, F. Wu, M. Trusiak, K. Patorski, Y. Wan, Q. Chen, and X. Hou, "Simultaneous extraction of phase and phase shift from two interferograms using Lissajous figure and ellipse fitting technology with Hilbert Huang prefiltering," J. Opt. **18**, 105604 (2016).
29. J. Vargas, J. A. Quiroga, and T. Belenguer, "Phase-shifting interferometry based on principal component analysis," Opt. Lett. **36**, 1326–1328 (2011).
30. K. Pearson. "On Lines and Planes of Closest Fit to Systems of Points in Space". Philosophical Magazine. 2 (11): 559–572 (1901).
31. J. Vargas, J. A. Quiroga, and T. Belenguer, "Analysis of the principal component algorithm in phase-shifting interferometry," Opt. Lett. **36**, 2215–2217 (2011).
32. J. Vargas, C. Sorzano, J. Estrada, and J. Carazo, "Generalization of the principal component analysis algorithm for interferometry," Opt. Commun. **286**, 130–134 (2013)..
33. J. Vargas and C. Sorzano, "Quadrature component analysis for interferometry," Opt. Lasers Eng. **51**, 637–641 (2013).
34. J. Xu, W. Jin, L. Chai, and Q. Xu, "Phase extraction from randomly phase-shifted interferograms by combining principal component analysis and least squares method," Opt. Express **19**, 20483–20492 (2011).
35. J. Deng, K. Wang, D. Wu, X. Lv, C. Li, J. Hao, J. Qin, and W. Chen, "Advanced principal component analysis method for phase reconstruction," Opt. Express **23**, 12222–12231 (2015).
36. K. Yatabe, K. Ishikawa, and Y. Oikawa, "Improving principal component analysis based phase extraction method for phase shifting interferometry by integrating spatial information," Opt. Express **24**, 22881–22891 (2016).
37. K. Yatabe, K. Ishikawa, and Y. Oikawa, " Simple, flexible, and accurate phase retrieval method for generalized phase-shifting interferometry," JOSA A. **34**, 6017–6024 (2016).
38. M. Servin, M. Padilla, G. Garnica and G. Paez, "Fourier spectra for nonuniform phase-shifting algorithms based on principal component analysis," Optics Express **27**, 25861-25871 (2019).
39. M. Servin, M. Padilla, G. Garnica, and G. Paez, "Design of nonlinear spaced phase-shifting algorithms using their frequency transfer function," Appl. Opt., **58**, 1134-1138 (2019).
40. D. Malacara, M. Servin and Z. Malacara, *Interferogram Analysis for Optical Testing*, 2nd. ed. (Taylor & Francis CRC 2005).
41. T. L. Schmitz, A. D. Davies, C. J. Evans, "Uncertainties in interferometric measurements of radius of curvature," Proc. SPIE **4451**, Optical Manufacturing and Testing IV, 432-447 (2001); doi: 10.1117/12.453641.